\documentclass{aa}
\usepackage{graphicx}
\usepackage{longtable}
\usepackage{lscape}

\input{psfig.tex}
\def\gsim{\;\lower.6ex\hbox{$\sim$}\kern-7.75pt\raise.65ex\hbox{$>$}\;}
\def\lsim{\;\lower.6ex\hbox{$\sim$}\kern-7.75pt\raise.65ex\hbox{$<$}\;}
\def\teff{T$_{\rm eff}$}
\def\ebv{E$(B-V)$}
\def\B{$B$}
\def\V{$V$}
\def\I{$I$}

\def\BV{$B-V$}
\def\VK{$V-K$}

\begin{document}
\title{High-resolution spectroscopy of the old 
open cluster Collinder 261: abundances of Iron and other elements
\thanks{
Based on observations collected at ESO telescopes under programme 65.N-0286}
 }

\author{
E. Carretta\inst{1},
A. Bragaglia\inst{1},
R.G. Gratton\inst{2},
\and
M. Tosi\inst{1}
}

\authorrunning{E. Carretta et al.}
\titlerunning{Abundances in Cr 261}

\offprints{E. Carretta, carretta@pd.astro.it}

\institute{
INAF - Osservatorio Astronomico di Bologna, Via Ranzani 1, I-40127
 Bologna, Italy
\and
INAF - Osservatorio Astronomico di Padova, Vicolo dell'Osservatorio 5, I-35122
 Padova, Italy
  }

\date{Received ; accepted }

\abstract{
We present the
analysis of high resolution spectra of six red giant stars
in the old open cluster Collinder 261. 
Reddening values for individual stars, derived
from the relation between colours and temperatures (deduced
from our fully spectroscopic analysis) are consistent with previous 
determinations based on photometry. 
For this cluster we derive an iron abundance of [Fe/H] = $-$0.03 $\pm$ 0.03. 
We also obtain the  abundances of light metals (O, Na and Al), $\alpha$-elements (Mg,
Si, Ca, Ti), elements of the Fe-group (Sc, Cr, Mn, Co, Ni) and the
neutron-capture element Ba. No intrinsic star-to-star scatter is present
in any of these elements within our sample. 
We compare our findings with previous investigations on this cluster,
discussing in detail differences in analysis methods and results. 
 }

\maketitle

\section{Introduction}

This is the second paper of a series devoted to the study of detailed elemental
abundances of open clusters (OCs), with the final goal of deriving the time
evolution of abundances and of the radial abundance gradient in the Galactic
disk.

Collinder 261 (Cr 261) is a populous system located in the inner Galactic disk
($l=301^{\circ}.68, ~b=-5^{\circ}.53$), 
and is of particular interest for such studies
because it is
one of the oldest disk clusters (with an age of 5 to 11 Gyr, see
Janes \& Phelps 1994; Mazur et al. 1995, Gozzoli
et al. 1996, Carraro et al. 1998). Despite its old age, both photometric (ibidem)
and spectroscopic (Friel at el. 2002, Friel et al. 2003) studies estimate 
a solar or only slightly sub-solar metallicity. This cluster
is therefore one of the few examples of
old and rather metal rich simple stellar population, and represents an
important  test for stellar evolutionary models. It is moreover an important
test particle to understand the Galaxy conditions at early epochs in the disk,
the effect of the environment on the survival and evolution of a bound system,
the effect of Galactic chemical evolution on the metallicity of clusters
formed at different distances from the Galactic centre, and the evolution of
the chemical abundance gradient.

We have been studying open clusters as tracers of the Galaxy evolution
for several
years, and to this purpose we are building a large and homogeneous
database of 
photometry and high-resolution spectroscopy, to infer ages, distances,
reddening values and chemical abundances. The most recent results from the
photometric project can be found in Tosi et al. (2004) and Bragaglia \& Tosi
(2005, and in preparation). Using high resolution spectroscopy of clump and red giant branch (RGB)
stars, we have analysed so far NGC 6819 (Bragaglia et al. 2001), NGC 2506, NGC
6134 and IC 4651 (Carretta et al. 2004a, hereafter C04). 

The metallicity of large samples of open clusters has been traditionally
derived from photometric data or  low-resolution spectroscopy (see e.g. Janes
1979, Panagia \& Tosi 1981, and the extensive analyses by Friel 1995,
Twarog et al. 1997, and Friel et al. 2002). Both these
methodologies lead to relatively inaccurate results and do not allow to
estimate the abundances of individual chemical species. So far, only in a
few cases high-resolution spectra  have been used for open clusters.  
Cr 261 is
one of them, since Friel et al. (2003, hereafter F03) have recently
derived abundances of iron, oxygen and other elements from high-resolution
spectra of four of its red giants. 
However, since {\em homogeneity} is one of the requirements of our global 
study of OCs, we have performed an independent study, applying our standard
abundance analysis to six stars  of Cr 261, using spectra with better resolution
than F03's.  Four of the stars are also  F03 targets, and have therefore
been used to evaluate the uncertainties on the abundances resulting from
different approaches, a very useful by-product.

The outline of this paper is the following: we describe our observations in
Section 2; we derive the atmospheric parameters and iron abundances in
Section 3, and we compare the reddening derived from our spectroscopy 
with the photometric one in Section 4; our results on abundances are presented
and extensively compared with the values inferred by F03
in Section 5, and with other clusters in Section 6; the overall results are 
summarized in Section 7.


\begin{figure}
\includegraphics[bb=100 200 500 600, clip, scale=0.6]{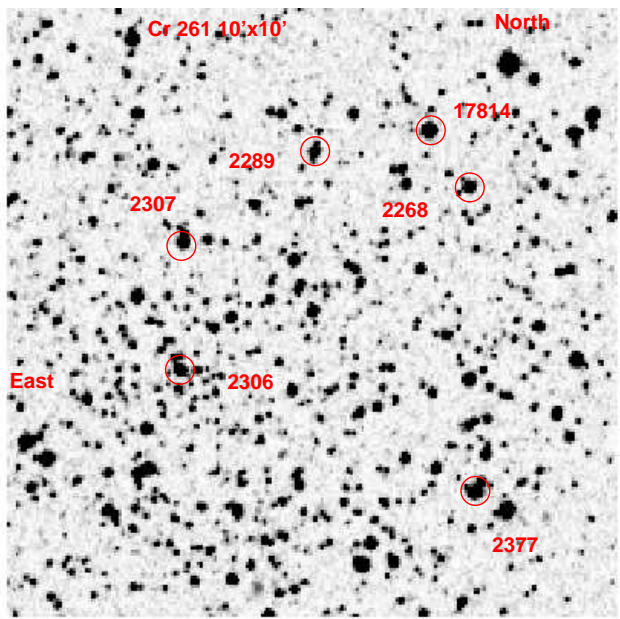}
\caption[]{Field of 10 $\times$ 10 arcmin$^2$ centered on Cr 261, with the 6
stars observed with FEROS indicated by their numbers according to Gozzoli
et al. (1996)}
\label{f:field}
\end{figure}

\section{Observations}

\begin{figure}
\includegraphics[bb=50 180 420 600, clip, scale=0.6]{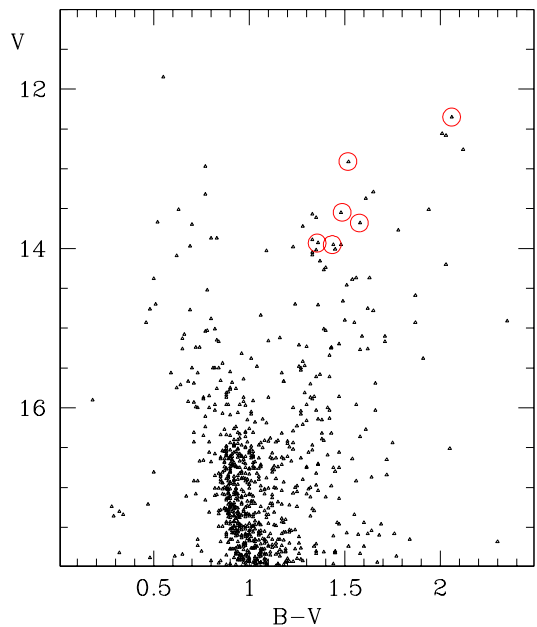}
\caption[]{CMD for Cr 261; the stars here analyzed are encircled.}
\label{f:cmd1}
\end{figure}

We observed six stars in Cr 261 with FEROS at the 1.5m telescope in La Silla
(Chile) from April 28 to May 1 2000 (see C04 for details). 
We selected  Red Clump (RC) and RGB stars from the photometric data of Gozzoli
et al. (1996) for which  membership information was kindly made available by E.
Friel in advance of publication. Core Helium burning RC stars are the best 
targets for spectroscopic studies of evolved populations, since they represent 
a quite homogeneous group, and their temperatures are sufficiently high 
for a safe analysis based on model atmospheres.
However, given the telescope size, the relatively faint magnitude of the RC in
Cr 261, and the need of high S/N spectra, we only observed two clump stars,
together with four RGB ones, two of which are only slightly brighter than
the clump.  A finding chart for all the observed targets is shown in
Fig.~\ref{f:field}, and the cluster CMD is presented in Fig. ~\ref{f:cmd1},
where the 6 stars are marked.  Table \ref{t:reldata1} gives a log of the
observations, and useful information on the selected stars.  S/N values are
measured  near 6700 \AA. For multiple exposures, radial velocity and S/N refer
to the final, co-added spectra.

\begin{table*}
\begin{center}
\caption{Log of the observations and relevant data for the target stars.
ID, in the first column, is the identification number used in Gozzoli et al. 
(1996), while the second column gives the ID
from Phelps et al. (1994) also used in
Friel et al. (2002, 2003).
Coordinates are at J2000 (in units of h:m:s, d:m:s);
$V$ and $B-V$ are taken from
Gozzoli et al. (1996) and K is the 2MASS value. S/N is computed near
$\lambda$=6700 \AA ~(on the summed spectrum), the radial velocity is heliocentric, 
and the exposure time is in seconds.}
\begin{tabular}{rrcccccrcccc}
\hline
 ID    &ID     & Ra      & Dec         & V      & B-V  & K        & S/N & RV         & Date obs.   & UT   &Exp \\ 
       &PJM    & (2000)  & (2000)      &        &      & 2MASS    &     & kms$^{-1}$ &             &start & (s)\\ 
\hline
17814  &1871 &12:37:43.608 &-68:19:55.06 &12.350  &2.060 & 7.907 & 130 & -26.85 & 2000 Apr 29 & 3:27:25 & 3600 \\ 
 2377  &2105 &12:37:34.828 &-68:23:24.99 &12.908  &1.517 & 9.184 &  95 & -25.34 & 2000 Apr 30 & 6:09:47 & 3600 \\ 
       &     &     	   &		 &	  &	 &	 &     &	& 2000 Apr 30 & 6:09:47 & 3600 \\ 
 2307  &1045 &12:38:08.499 &-68:21:15.01 &13.547  &1.487 & 9.788 &  85 & -25.56 & 2000 May 01 & 3:16:43 & 3600 \\ 
       &     &     	   &		 &	  &	 &	 &     &	& 2000 May 01 & 3:16:43 & 3600 \\ 
 2289  &1485 &12:37:55.557 &-68:20:14.36 &13.680  &1.577 & 9.801 &  85 & -27.63 & 2000 Apr 29 & 4:37:56 & 3600 \\ 
       &     &     	   &		 &	  &	 &	 &     &	& 2000 Apr 29 & 5:41:30 & 3600 \\ 
 2268  &2001 &12:37:38.681 &-68:20:25.87 &13.932  &1.356 &10.696 &  75 & -25.37 & 2000 Apr 29 & 6:48:44 & 3600 \\ 
       &     &     	   &		 &	  &	 &	 &     &	& 2000 Apr 30 & 3:45:22 & 3600 \\ 
       &     &     	   &		 &	  &	 &	 &     &	& 2000 Apr 30 & 4:48:33 & 3600 \\ 
 2306  &1080 &12:38:07.420 &-68:22:30.82 &13.952  &1.435 &10.343 &  70 & -25.31 & 2000 May 01 & 5:42:01 & 3600 \\ 
       &     &             &             &        &      &       &     &        & 2000 May 01 & 6:45:03 & 3600 \\ 
\hline

\hline
\end{tabular}
\label{t:reldata1}
\end{center}
\end{table*}


\section{Analysis}

\subsection{Atmospheric parameters}

To 
derive effective temperatures, gravities and microturbulent velocities
we followed exactly the procedure described in C04 (Section 3).
Our adopted atmospheric parameters, derived from a
fully spectroscopic analysis, are listed in Table~\ref{t:atmpar}.

\begin{table*}
\begin{center}
\caption[]{Adopted atmospheric parameters and derived Iron abundances; nr 
indicates the number of lines used in the analysis.}
\begin{tabular}{rcccrcccrcc}
\hline
Star   &  T$_{\rm eff}$ & $\log$ $g$ & $v_t$         & nr & log(n)      & $rms$ & [Fe/H]{\sc i}& nr & log(n)       & $rms$ \\
       &     (K)        &  (dex)     & (km s$^{-1}$) &   & Fe {\sc i}  &       &              &   & Fe {\sc ii}  &        \\
\hline
17814 & 3980 & 0.43  &1.44 & 94 &7.223 &0.139 &-0.31 &  9 &7.182 &0.190 \\ 
 2377 & 4180 & 1.59  &1.29 &124 &7.456 &0.145 &-0.08 &  8 &7.408 &0.188 \\ 
 2307 & 4470 & 2.07  &1.23 &126 &7.548 &0.157 &~0.00 & 10 &7.493 &0.189 \\ 
 2289 & 4340 & 1.76  &1.27 &113 &7.484 &0.128 &-0.06 & 12 &7.433 &0.172 \\ 
 2268 & 4580 & 1.83  &1.26 &122 &7.522 &0.114 &-0.02 & 11 &7.472 &0.188 \\ 
 2306 & 4500 & 2.09  &1.23 &114 &7.536 &0.152 &~0.00 & 11 &7.485 &0.256 \\ 
\hline
\end{tabular}
\label{t:atmpar}
\end{center}
\end{table*}

\subsection{Equivalent widths}

We measured the $EW$s  employing an  updated version of  the
spectrum analysis package developed in Padova by one of us (RG), 
following the method described
in Bragaglia et al. (2001). Continuum tracing, a difficult task at these
metallicities and cool temperatures, was checked with spectrum synthesis  of Fe
{\sc i} and Fe {\sc ii} lines, and the ensuing analysis was done as amply
described in C04.

Fe lines were considered only in the interval 5500-7000~\AA\ to minimize
problems in the continuum tracing due to line crowding blueward of this region, and
contamination by telluric lines and possible fringing effects in the red.
Sources of oscillator strengths and atomic parameters are the same as 
given by Gratton et al. (2003): discussion and references  can be found in that
paper.

Errors in $EW$s are estimated to be 4 m\AA, and were determined comparing the
$EW$s of Fe {\sc i} lines in the two stars 2307 and 2268, of similar
atmospheric parameters and  
intermediate S/N ratios.

\begin{table*}
\begin{center}
\caption[]{Sensitivities of abundance ratios to variations in the atmospheric
parameters and in the equivalent widths, as computed for the RC star
2268. The total error on [Fe/H]{\sc ii} is not given, as Fe {\sc ii} was forced 
to agree with Fe {\sc i}. The total error is computed as the sum 
of the two dominant sources of error,
\teff \ and $v_t$ (Col. 8: tot.1) or as the sum of all
contributions (Col. 9: tot.2), as described in the text.}
\begin{tabular}{lrrrrrrrr}
\hline
\\
Ratio    & $\Delta T_{eff}$ & $\Delta$ $\log g$ & $\Delta$ [A/H] & $\Delta v_t$
&$<N_{lines}>$& $\Delta$ EW & tot.1 & tot.2\\
         & (+60 K)    & (+0.2 dex)      & (+0.05 dex)      & (+0.14 km/s) & & & (dex)& (dex)  \\
 (1) & (2) & (3) & (4) & (5) & (6) & (7) & (8) & (9) \\
\\
\hline
$[$Fe/H$]$I  &  +0.037 &   +0.017 &  +0.006 & $-$0.038 &120 &+0.013  &0.053 &0.058  \\
$[$Fe/H$]$II &$-$0.067 &   +0.111 &  +0.018 & $-$0.028 & 10 &+0.044  &      &	    \\
\hline
$[$O/Fe$]$I  &  +0.001 & $-$0.019 &$-$0.012 &    0.069 &  4 &+0.070  &0.069 &0.101  \\
$[$Na/Fe$]$I &  +0.018 & $-$0.058 &$-$0.005 &    0.046 &  6 &+0.057  &0.049 &0.095  \\
$[$Mg/Fe$]$I &$-$0.005 & $-$0.040 &$-$0.001 &    0.053 &  8 &+0.049  &0.053 &0.083  \\
$[$Al/Fe$]$I &  +0.010 & $-$0.022 &$-$0.009 &    0.064 &  2 &+0.099  &0.065 &0.121  \\
$[$Si/Fe$]$I &$-$0.061 &   +0.018 &  +0.004 &    0.059 & 17 &+0.034  &0.085 &0.093  \\
$[$Ca/Fe$]$I &  +0.032 & $-$0.047 &$-$0.006 &    0.011 & 17 &+0.034  &0.034 &0.067  \\
$[$Sc/Fe$]$II&  +0.062 & $-$0.031 &$-$0.002 & $-$0.022 &  9 &+0.047  &0.066 &0.087  \\
$[$Ti/Fe$]$I &  +0.063 & $-$0.025 &$-$0.009 & $-$0.008 & 23 &+0.029  &0.064 &0.075  \\
$[$Ti/Fe$]$II&  +0.060 & $-$0.033 &$-$0.001 & $-$0.010 & 10 &+0.044  &0.061 &0.082  \\
$[$Cr/Fe$]$I &  +0.033 & $-$0.029 &$-$0.006 &    0.008 & 38 &+0.023  &0.034 &0.051  \\
$[$Cr/Fe$]$II&  +0.022 & $-$0.026 &$-$0.005 &    0.021 &  8 &+0.049  &0.030 &0.063  \\
$[$Mn/Fe$]$I &  +0.037 & $-$0.034 &  +0.001 &    0.002 &  7 &+0.053  &0.037 &0.073  \\
$[$Co/Fe$]$I &$-$0.007 &   +0.014 &  +0.002 &    0.016 &  8 &+0.049  &0.017 &0.054  \\
$[$Ni/Fe$]$I &$-$0.020 &   +0.020 &  +0.003 &    0.023 & 37 &+0.023  &0.030 &0.043  \\
$[$Ba/Fe$]$II&  +0.080 & $-$0.045 &  +0.002 & $-$0.051 &  2 &+0.099  &0.095 &0.144  \\
\hline
\end{tabular}
\label{t:sensitivity}
\end{center}
\end{table*}

\subsection{Errors on atmospheric parameters}

Uncertainties in atmospheric parameters have been estimated as in C04. We find
values very similar to the ones in C04, as expected since we are dealing with
stars of similar metallicity and evolutionary status observed with the same
instrumentation. We give here only a concise presentation, and refer to C04 for
an extensive description and discussion. 

\paragraph{Errors in effective temperatures:} 
they were estimated from the errors in the slope of the relation between Fe
{\sc i} abundances and line excitation potentials.
If we exclude the star near the tip  (in C04 we found that the adopted model
atmospheres from the Kurucz grid are not adequate for the analysis of stars
near the tip of the Red Giant Branch of open clusters), we estimate a standard
error of 93 $\pm$ 21 (rms=47) K; we adopt 90 K, that corresponds to an average
rms of 0.013 dex eV$^{-1}$ in the slope.

This error includes two different terms: i) a random component (the true
internal error), that affects $EW$s measurement, and that reflects errors
from Poisson statistics, read-out noise, etc; and ii) a systematic
component, that is the same for each line in all stars, and reflects e.g.
the presence of blends, or uncertain line oscillator strengths. 
As computed in C04, the random
component is about 60 \% of the total error, so the internal random error
in \teff \ is  56 K.

\paragraph{Errors in surface gravities:} 
 since our parameters have been derived entirely from  spectroscopy, there
are only  two contributions to random errors: from the uncertainty in \teff \,
and from the measurement errors of the individual lines. To determine the
first component, one has to take into account the variation in ionization
equilibrium due to changes in atmospheric parameters. From Table
\ref{t:sensitivity} (see below for a description),  considering our random
error in \teff, we derive a contribution to the random error in $\log g$ of
0.22 dex.

The second component may be evaluated weighting the average rms of abundances
from a single line (random error: 0.088 dex) with the number of lines (120 for
Fe {\sc i}  and 10 for Fe {\sc ii}) and finding, using again Table
\ref{t:sensitivity}, what difference in gravity is implied. We
find that the $EW$s contribute 0.067 dex to the random error in $\log g$.

Adding in quadrature the two components, we obtain a total random uncertainty
in the adopted gravity of 0.23 dex.

\paragraph{Errors in microturbulent velocities:} 

We used star 2289 and repeated the analysis changing $v_t$ until the 1 $\sigma$
value from the original slope of the relation line strengths-abundances was
reached; the corresponding error is 0.15 km s$^{-1}$, 60 \% of which is random,
i.e. 0.09 km s$^{-1}$.

Since the $v_t$ values are derived from a relation (see C04), we have to take 
into account also the rms scatter (0.17 km s$^{-1}$) around this relation.  To
obtain the final estimate of the error we have to subtract (in quadrature) from
this value the random error given just above. The final error on $v_t$ is then
0.14  km s$^{-1}$.

\paragraph{Sensitivity of abundances to atmospheric parameters:}
Table \ref{t:sensitivity} shows the sensitivity of the derived abundances to
variations in atmospheric parameters (Cols. 2 to 5; note that these variations
have the same size of the actual internal errors), and to errors in $EW$s
(Col. 7, where the average error from a single line is weighted by the square
root of the mean number of lines, given in Col. 6). This is done for iron and
for the other elements measured in this paper, and is derived for the RC
star 2268, as representative of the whole sample.

As discussed in C04, these sensitivities are computed assuming that
errors on the
single parameters are independent of each other, but this is not the case:
e.g., \teff's and gravities, and \teff's and microturbulent velocities are
strictly correlated. Col. 8 of Table \ref{t:sensitivity} shows the random
error in abundance for all the elements considered, taking into account only
the dominant uncertainty sources, i.e. \teff \ and $v_t$ and the effect of
their covariance. This amounts to 0.053 dex for [Fe/H]{\sc i}.  Taking into
account all the uncertainty sources, we obtain instead 0.058 dex. This second kind
of estimate is given in Col. 9, also for all other elements.

\section{Reddening estimates from spectroscopy}

Our \teff's are completely spectroscopic, i.e. reddening-free, so they can
be used to derive an estimate of the cluster reddening independently of the
photometric determination.
The method has been discussed in C04. Shortly, from
the \teff ~and the colour-temperature relation of Alonso et al.
(1999) we determine the de-reddened colour for each star, and the reddening 
follows from the comparison with the observed colour. 

This can be done for several photometric
systems;  we have searched the BDA (Mermilliod 1995, and {\tt
http://obswww.unige.ch/webda/webda.html}) and the Simbad databases  for data on
Cr 261.
There are two papers presenting \B ~and \V ~photometry,
namely Gozzoli et al. (1996)
and Mazur et al. (1995; but they have published information only for the
variable stars). Janes et al. (1994) provide only \V ~and \I; no data in the
Str\"omgren system was found. We took the  $JHK$ photometry from 2MASS (Cutri
et al. 2003: Point Source Catalogue, All-Sky Data Release, {\tt
http://www.ipac.caltech.edu/2mass/}) and transformed it to the TCS system. We
adopted the colour-temperature transformations by Alonso et al. (1999) for \BV
~and \VK, taking into account the cluster metallicity,  and derived an estimate
of the \ebv ~value [adopting  $E(V-K)=2.75 E(B-V)$: Cardelli et al. 1989] for
all stars except the one near the RGB tip, which is too cool. 
Results are presented in Table ~\ref{t:reddenoi}, where the average reddening
values
were computed without the two doubtful cases (see below).

\begin{table}
\begin{center}
\caption{Values for the individual reddening values derived from B--V (col. 2)
and V--K (col. 3). The average values are computed without the uncertain
values, indicated by a ":".}
\begin{tabular}{cccl}
\hline
 Star   & E(B-V) & E(B-V)& Notes\\
        & (B-V)  & (V-K) & \\
\hline
 2377   & 0.150:  & 0.222: & contam. ?\\
 2307   & 0.307  & 0.397   & \\
 2289   & 0.316  & 0.374   & \\  
 2268   & 0.238  & 0.254:  & contam. ?\\ 
 2306   & 0.274  & 0.358   & \\
\hline
average & 0.299  & 0.376   & 3 stars\\
$\sigma$& 0.022  & 0.020   & \\    
\hline
\end{tabular}
\label{t:reddenoi}
\end{center}
\end{table}

\begin{figure}
\includegraphics[bb=30 190 358 420, clip, scale=0.75]{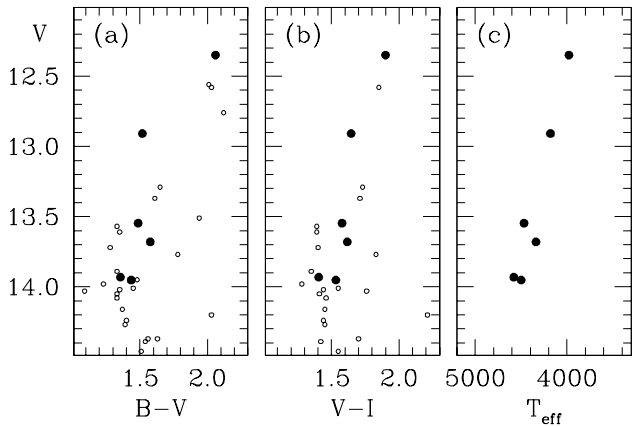}
\caption[]{(a) V, B-V CMD for the RGB and RC region of Cr 261, with the 
stars analyzed here indicated by larger symbols; (b) the same, for the V,V-I
CMD; (c) only the observed stars are shown in the V-\teff ~diagram}
\label{f:cmd2}
\end{figure}

We notice that there are two discrepant cases (star 2377 both in \BV ~and \VK, 
and star 2268 only in \VK), with reddening values much lower than the other
stars. Excluding a strong
differential reddening, which was not found by any photometric study, 
a very plausible explanation is that 
their bluer colours are caused by
blends;
being these stars
among the reddest ones in the cluster, the possibility of a blend with a bluer
object is not negligible.

We have checked our original photometric images for star 2377, but could not
find any strong indication in favour of (or against) another component. Star
2377 is about 0.12 mag bluer than expected, with  \BV ~= 1.52 instead of 1.64.
We have estimated the effect of possible blends  on  star 2377, 
and found that we would get B--V and V--K
colours consistent with the observed ones by
blending an RGB star of the same magnitude as star
2377 with an object just brighter than the MS turn-off of Cr 261 (i.e. with
V$\simeq$16.0, B$\simeq$16.9 and K$\simeq$14.8). Since such contaminating
candidates are numerous in the observed field, the chances for blends
are not negligible. Moreover, we cannot exclude additional blends in the K band,
due to the worse resolution of the 2MASS photometry.  The
faintness of the blended objects relative to our RGB stars explains the absence
of related features in the observed spectra.

Furthermore, star 2377 has an anomalous position in the $V$,\BV ~CMD
(see Fig. ~\ref{f:cmd2}(a)), being bluer than the expected RGB locus, but 
its position is closer to the RGB in the \V,$V-I$ 
~CMD (Fig.~\ref{f:cmd2}(b)), and absolutely correct if we consider its \teff 
\ (Fig. ~\ref{f:cmd2}(c)).  We take this
as further evidence in favour of a blend.  

A second interesting result is that 
we derive from \BV ~a reddening lower by 0.077 mag than from \VK: 
0.299 ($\sigma$=0.022) and 0.376 ($\sigma$=0.020) respectively. 
This may require a value of R, the ratio between selective and total absorption,
higher than the 'normal' 3.1: a value of R near
3.8 would be more appropriate to explain the difference. This value is not 
extreme (see e.g. the case of M 4, where Ivans et al. (1999) find R = 3.4
$\pm$ 0.4). Geminale \& Popowski (2004) have recently determined
R$_V$ values for Milky Way stars using ultraviolet colour excesses, and the 
ones for the two
stars nearest to Cr 261 are definitely larger than 3.1: 
HD109399, with an approximate distance to  Cr 261 of 4.4 deg has R$_V$ = 4.26,
while HD109867, with  a distance of about  1.1 deg, has R$_V$ = 3.67.

Another solution would be an error in the photometry, either in the 2MASS
values, or in the Johnson-Cousins ones of Gozzoli et al. (1996). In the latter
case, this  would require an error of about 0.08 mag in \BV, which
is hard to justify: Gozzoli et al. (1996) compared their CMDs
to the ones by Mazur et al. (1995) and found them perfectly consistent, even
if no star-to-star comparison could be done.

Of course, a combination of both explanations is also possible: new photometry
or larger samples of spectra are required to settle this matter.

Finally, we notice that the infrared maps of Schlegel et al. (1998) give \ebv =
0.433 for the location of the cluster; however Cr 261 is near the Galactic
plane ($b = -5^{\circ}.53$) and Schlegel et al. caution against use of
their maps for $|b| \le 5^{\circ}$.

\section{Comparison with previous works: metallicity}

Cr 261 was included among the old OCs studied by Friel et al. (2002)
by means of moderate resolution spectroscopy (1.8 \AA ~pix$^{-1}$)
with Argus at CTIO.
They observed 25 bright stars, confirming membership for 21 of them, 
and deduced a mean metallicity of [Fe/H]=$-$0.16 dex ($\sigma$=0.13 dex).

F03 observed stars 17814, 2377, 2307, and 2306 at a resolution of R =
25000, and typical S/N of 80, using the echelle spectrograph mounted at the
CTIO 4m telescope. Our spectra have similar S/N and 
exposure times, but about twice their resolution; this is due to the very high
efficiency of FEROS compared to the spectrograph used by F03.

To overcome problems with line-blanketing, which is
rather severe in these cool and metal
rich stars, F03 determined the continuum level of their spectra from the
comparison with a spectrum of Arcturus of similar resolution. 
From the analysis of
equivalent widths they estimated an average iron abundance for the cluster of
[Fe/H] = $-0.22 \pm$ 0.05, and roughly solar oxygen and $\alpha$-elements
abundance ratios.

\begin{figure}
\includegraphics[bb=35 190 380 410, clip, scale=0.7]{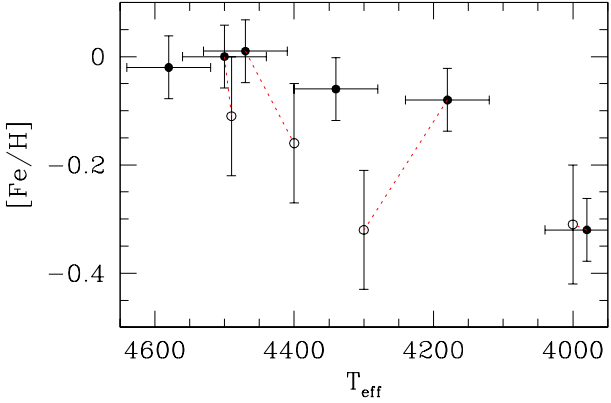}
\caption[]{[Fe/H] values derived in this paper (filled symbols) and
by F03 (open symbols) plotted against \teff; lines connect
the values for the 4 stars in common. Error bars on our values come from Table
3.}
\label{f:friel1}
\end{figure}

Table \ref{noifriel} presents a direct comparison of values obtained in the
present work and by F03 for radial velocity, atmospheric parameters, and iron
abundance.  Fig. ~\ref{f:friel1} shows a graphic comparison between
abundances of Fe {\sc i}, plotted against \teff. A trend of lower [Fe/H] with
decreasing \teff \ is clearly visible; this is not unusual for analyses of
cool, metal rich stars. Notice however that in our study the five stars with
higher  \teff \ give [Fe/H] values in complete agreement with each other.
The only really discrepant value is for the star near the RGB tip (for
which, nonetheless, F03 results are almost identical to ours).

\begin{table*}\begin{center}
\caption{Comparison between our results (columns 1 to 6) and those in F03
(columns 7 to 12).}
\begin{tabular}{rrrrrrrrrrrr}
\hline  
 \multicolumn{1}{c}{ID}
&\multicolumn{1}{c}{RV}
&\multicolumn{1}{c}{T$_{\rm eff}$}
&\multicolumn{1}{c}{$\log g$}
&\multicolumn{1}{c}{$v_t$}
&\multicolumn{1}{r}{${\rm [Fe/H]}$}      				
&\multicolumn{1}{c}{ID$_{\rm F03}$}
&\multicolumn{1}{c}{RV$_{\rm F03}$}
&\multicolumn{1}{c}{T$_{\rm eff}$$_{\rm F03}$}
&\multicolumn{1}{c}{$\log g$$_{\rm F03}$}
&\multicolumn{1}{c}{$v_t$$_{\rm F03}$}
&\multicolumn{1}{r}{${\rm [Fe/H]}_{\rm F03}$}   \\  				
 \multicolumn{1}{c}{}
&\multicolumn{1}{c}{(km s$^{-1}$)}
&\multicolumn{1}{c}{(K)}
&\multicolumn{1}{c}{(dex)}
&\multicolumn{1}{c}{(km s$^{-1}$)}
&\multicolumn{1}{r}{}      				
&\multicolumn{1}{c}{}
&\multicolumn{1}{c}{(km s$^{-1}$)}
&\multicolumn{1}{c}{(K)}
&\multicolumn{1}{c}{(dex)}
&\multicolumn{1}{c}{(km s$^{-1}$)}
&\multicolumn{1}{r}{}		  \\				    
\hline        				
17814 &-26.85 &3980 &0.43 &1.44 &-0.32  &1871  &-26.0 &4000 &0.7  &1.5 & -0.31  \\ 
 2377 &-25.34 &4180 &1.59 &1.29 &-0.08  &2105  &-25.2 &4300 &1.5  &1.5 & -0.32  \\ 
 2307 &-25.56 &4470 &2.07 &1.29 & 0.01  &1045  &-24.8 &4400 &1.5  &1.2 & -0.16  \\ 
 2306 &-25.31 &4500 &2.09 &1.26 & 0.00  &1080  &-24.5 &4490 &2.2  &1.2 & -0.11  \\
\hline  
\label{noifriel}      				
\end{tabular}\end{center}\end{table*}

\begin{figure*}
\includegraphics[bb=20 170 555 420, clip, scale=0.9]{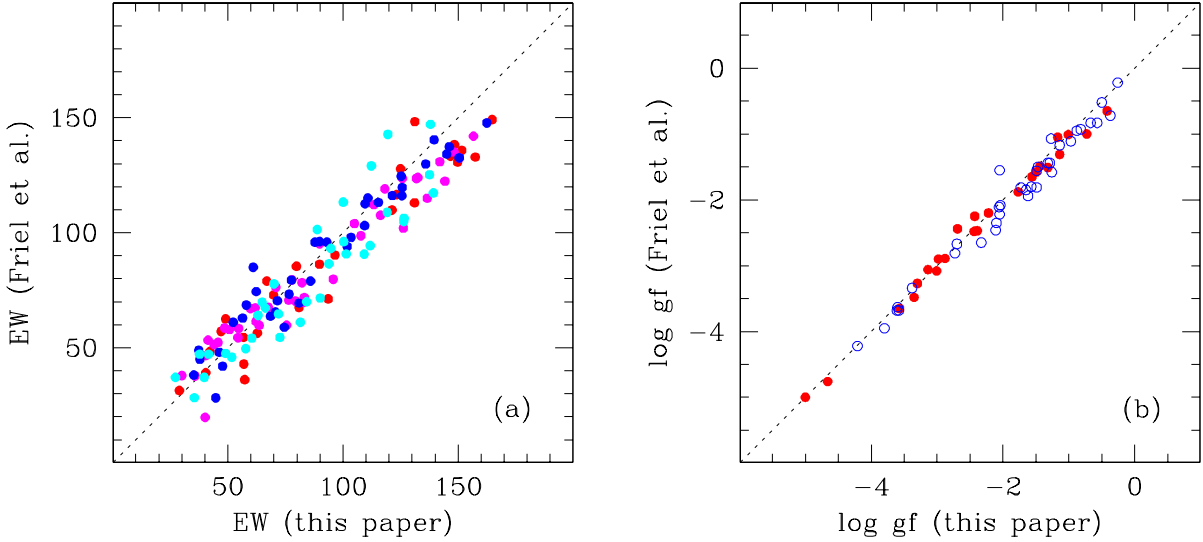}
\caption[]{(a) Comparison of the $EW$s of lines in common between our work and F03;
the mean difference (us -- F03)  is 3.8 m\AA, with rms = 10.7 m\AA. (b)
Comparison of log$gf$s used for iron lines (filled dots) and other elements
(open circles); the mean difference (us -- F03), considering only the 24 Fe
I lines in common used in our abundance analysis, is 0.04, with rms = 0.12.}
\label{f:conf}
\end{figure*}

Several factors may contribute to these differences:

i) The use of different atmospheric models (Kurucz and MARCS). However, both
analyses are differential (with respect to the Sun and Arcturus), and the 
difference in abundances should be minimal.

ii) The different format and resolution of the spectra. We have a larger
coverage in wavelength, so we measure more than 100 Fe {\sc i} lines for each
star, instead of about 40 as in the F03 study. Our resolution is about twice as
large, which allows easier
continuum tracing and $EW$ measurement of blended
lines. We have compared the $EW$s for the lines in common [see
Fig.\ref{f:conf}(a)]: the mean difference between the two sets of $EW$s  is
3.8 m\AA \  (rms = 10.7 m\AA), in  the sense ours minus F03's. The
internal error on $EW$s is 4 m\AA \ in our spectra,  and 7 m\AA \ in F03. This
translates also into a different dispersion in the Fe abundances, which is less than
0.14 dex in our case and slightly more than 0.2 dex in F03.

iii) Different $gf$'s. The two studies adopt slightly different $gf$ values [as
shown if Fig. \ref{f:conf}(b)], ours being on average larger\footnote{
F03 compared their  $gf$'s with literature sources, finding that their
values were systematically
smaller}. 
For the 24 Fe {\sc i} lines in common, the average difference is 0.04
dex. This implies the derivation of larger values for the microturbulent
velocity $v_t$'s and smaller abundances.

iv) Other differences in the analysis methods. These effects are
more difficult to assess, given the
number of assumptions and methodologies.   There is one point that
surely introduces large differences: the determination of the
microturbulent velocity by zeroing the slope of abundances from individual Fe
{\sc i} lines versus $EW$s. As we understand it, F03 found $v_t$ eliminating
trends with {\em measured} $EW$s, while we used a mean relation between $v_t$
and $\log g$. The values of $v_t$ used to this purpose were found by zeroing
the trend of Fe {\sc i} abundances versus   
$EW$s {\em expected}  on the basis of model atmospheres.  Our
approach follows the prescriptions by Magain (1984),  and it should be
preferred to the technique used by F03 because in the latter  case random
errors on measured $EW$s produce systematically larger $v_t$'s, and 
hence smaller
abundances. To quantify this effect, we have re-determined the Fe {\sc i}
abundance for star 2377, the one for which the [Fe/H] values  differ most,
applying our method to F03 $EW$s and $gf$'s, and with Kurucz atmospheres.   We
started from the published atmospheric parameters and iterated until no trend
with expected $EW$s was present; the final solution we found for this star is
\teff=4320, $\log g$=1.60, $v_t$=1.25, and [Fe/H]=$-$0.07, much closer to our
values.   We suggest that the different approach
to estimate $v_t$ is  the major source of the 
abundance difference.
  
The adopted solar abundances for iron are very close in the two studies: 
$\log n$(Fe)=7.54 in our
case  and 7.52 in F03 (E. Friel, private
communication). If we translate their results for the metal abundance of
Cr 261 to our scale, we  obtain [Fe/H]=$-$0.24 dex, slightly increasing
the nominal difference.

\begin{table}
\begin{center}
\caption[]{Abundances of all the elements measured in the six stars. The
C abundance and the first value for the O abundance (indicated with "a") 
are preliminary estimates based on complete synthesis, while the second value
for O (indicated with "b") comes
from the $EW$ of the 6300.31 \AA \ line. Sc, V, Mn and Co have HFS taken
into account; Na value is in  N-LTE. The last column gives the Solar values adopted
in this analysis.}
{\scriptsize
\begin{tabular}{l rrr rrr r}
\hline
element & nr  & abu. &$\sigma$ & nr  & abu. &$\sigma$ &  Solar\\
\hline
        &\multicolumn{3}{c}{---------17814---------} &\multicolumn{3}{c}{---------2289---------} &\\
${\rm [Fe/H]}${\sc i}    & 94 &$-$0.32 &0.14 &113 &$-$0.06 &0.13  & 7.54\\
${\rm [Fe/H]}${\sc ii}   &  9 &$-$0.31 &0.19 & 12 &$-$0.06 &0.17  & 7.49\\ 
${\rm [C/Fe]}${\sc i} a  &    &        &     &  1 &$-$0.08 &	  & 8.52\\
${\rm [O/Fe]}${\sc i} b  &  1 &$-$0.57 &     &  1 &$-$0.25 &	  & 8.79\\
${\rm [O/Fe]}${\sc i} a  &    &        &     &  1 &$-$0.04 &	  & 8.79\\
${\rm [Na/Fe]}${\sc i}   &  5 & 0.87   &0.17 &  5 & 0.36   &0.08  & 6.21\\ 
${\rm [Mg/Fe]}${\sc i}   &  8 & 0.31   &0.15 &  5 & 0.09   &0.17  & 7.43\\ 
${\rm [Al/Fe]}${\sc i}   &  5 & 0.37   &0.13 &  6 & 0.18   & 0.19 & 6.23\\
${\rm [Si/Fe]}${\sc i}   &  9 & 0.13   &0.11 & 18 & 0.22   & 0.21 & 7.53\\
${\rm [Ca/Fe]}${\sc i}   & 11 & 0.29   &0.10 & 16 & 0.08   & 0.13 & 6.27\\
${\rm [Sc/Fe]}${\sc ii}  &  5 &$-$0.19 &0.17 &  6 & 0.16   &0.14  & 3.13\\
${\rm [Ti/Fe]}${\sc i}   & 24 & 0.28   &0.23 & 23 &$-$0.06 &0.22  & 5.00\\
${\rm [Ti/Fe]}${\sc ii}  & 11 &$-$0.03 &0.21 & 12 &$-$0.11 &0.15  & 5.07\\
${\rm [Cr/Fe]}${\sc i}   & 33 &$-$0.02 &0.19 & 38 & 0.07   & 0.17 & 5.67\\
${\rm [Cr/Fe]}${\sc ii}  &  3 & 0.32   &0.04 &  7 &$-$0.08 & 0.12 & 5.71\\
${\rm [Mn/Fe]}${\sc i}   &  3 & 0.14   &0.07 &  6 &$-$0.02 & 0.16 & 5.34\\
${\rm [Co/Fe]}${\sc i}   &  4 &$-$0.10 &0.22 &  4 &$-$0.12 & 0.07 & 4.92\\
${\rm [Ni/Fe]}${\sc i}   & 28 & 0.16   &0.18 & 36 &$-$0.03 & 0.19 & 6.28\\
${\rm [Ba/Fe]}${\sc ii}  &  2 & 0.35   &0.06 &  2 & 0.35   & 0.04 & 2.22\\ 
\hline
        &\multicolumn{3}{c}{---------2268---------}  &\multicolumn{3}{c}{---------2377---------} & \\
${\rm [Fe/H]}${\sc i}   &122 &$-$0.02 &0.11   &124 &$-$0.08 &0.15  & 7.54\\
${\rm [Fe/H]}${\sc ii}  & 11 &$-$0.02 &0.19   &  8 &$-$0.08 &0.19  & 7.49\\
${\rm [C/Fe]}${\sc i} a &  1 &$-$0.16 &       &  1 &$-$0.07 &	   & 8.52\\
${\rm [O/Fe]}${\sc i} b &  1 &$-$0.30 &       &  1 &$-$0.35 &	   & 8.79\\
${\rm [O/Fe]}${\sc i} a &  1 &$-$0.23 &       &  1 &$-$0.08 &	   & 8.79\\
${\rm [Na/Fe]}${\sc i}  &  5 & 0.22   &0.03   &  5 & 0.37   &0.07  & 6.21\\
${\rm [Mg/Fe]}${\sc i}  &  8 & 0.19   &0.23   &  7 & 0.30   &0.19  & 7.43\\
${\rm [Al/Fe]}${\sc i}  &   6& 0.10   & 0.14  &   6& 0.22   &0.14  & 6.23\\
${\rm [Si/Fe]}${\sc i}  &  16& 0.16   & 0.12  &  16& 0.31   &0.18  & 7.53\\
${\rm [Ca/Fe]}${\sc i}  &  10&$-$0.01 & 0.09  &  9 &$-$0.04 &0.08  & 6.27\\
${\rm [Sc/Fe]}${\sc ii} &  6 & 0.06   &0.13   & 7  & 0.15   &0.14  & 3.13\\
${\rm [Ti/Fe]}${\sc i}  & 23 &$-$0.21 &0.20   & 23 &$-$0.15 &0.23  & 5.00\\
${\rm [Ti/Fe]}${\sc ii} &  9 &$-$0.11 &0.17   & 9  &$-$0.02 &0.11  & 5.07\\
${\rm [Cr/Fe]}${\sc i}  &  40&$-$0.06 & 0.19  &  36&$-$0.02 & 0.19 & 5.67\\
${\rm [Cr/Fe]}${\sc ii} &   7& 0.03   & 0.10  &   5& 0.19   & 0.24 & 5.71\\
${\rm [Mn/Fe]}${\sc i}  &   6& 0.04   & 0.16  &   6&$-$0.07 & 0.10 & 5.34\\
${\rm [Co/Fe]}${\sc i}  &   4& 0.11   & 0.21  &   5& 0.00   & 0.17 & 4.92\\
${\rm [Ni/Fe]}${\sc i}  &  44& 0.02   & 0.16  &  46& 0.12   & 0.19 & 6.28\\
${\rm [Ba/Fe]}${\sc ii} &   2& 0.15   & 0.09  &   2& 0.34   & 0.02 & 2.22\\
\hline
        &\multicolumn{3}{c}{---------2307---------}  &\multicolumn{3}{c}{---------2306---------} &\\
${\rm [Fe/H]}${\sc i}   &126 & 0.01   &0.16  &114 &$-$0.00 &0.15 & 7.54\\
${\rm [Fe/H]}${\sc ii}  & 10 & 0.00   &0.19  & 11 &$-$0.01 &0.26 & 7.49\\
${\rm [C/Fe]}${\sc i} a &  1 &$-$0.26 &       &  1 &  0.02 &	 & 8.52\\
${\rm [O/Fe]}${\sc i} b &  1 &$-$0.34 &      &  1 &$-$0.25 &	 & 8.79\\
${\rm [O/Fe]}${\sc i} a &  1 &$-$0.21 &      &  1 &$-$0.06 &	 & 8.79\\
${\rm [Na/Fe]}${\sc i}  &  5 & 0.38   &0.08  &  5 & 0.33   &0.09 & 6.21\\
${\rm [Mg/Fe]}${\sc i}  &  9 & 0.13   &0.16  &  8 & 0.15   &0.16 & 7.43\\
${\rm [Al/Fe]}${\sc i}  &  6 &$-$0.02 & 0.18 &  6 & 0.14   &0.11 & 6.23\\
${\rm [Si/Fe]}${\sc i}  & 16 & 0.26   & 0.19 & 17 & 0.23   &0.17 & 7.53\\
${\rm [Ca/Fe]}${\sc i}  & 14 &$-$0.04 & 0.16 & 0  & 0.04   &0.13 & 6.27\\
${\rm [Sc/Fe]}${\sc ii} &  6 & 0.15   & 0.16 &  6 & 0.10   &0.09 & 3.13\\
${\rm [Ti/Fe]}${\sc i}  & 21 &$-$0.19 & 0.21 & 21 & 0.03   &0.23 & 5.00\\
${\rm [Ti/Fe]}${\sc ii} & 12 &   0.00 & 0.16 &  8 &$-$0.03 &0.14 & 5.07\\
${\rm [Cr/Fe]}${\sc i}  & 34 &$-$0.10 & 0.15 & 40 & 0.05   &0.21 & 5.67\\
${\rm [Cr/Fe]}${\sc ii} &  6 & 0.14   & 0.10 &  5 & 0.05   &0.39 & 5.71\\
${\rm [Mn/Fe]}${\sc i}  &  6 &$-$0.06 & 0.17 &  6 &$-$0.05 &0.14 & 5.34\\
${\rm [Co/Fe]}${\sc i}  &  4 & 0.13   & 0.05 &  5 &$-$0.06 &0.19 & 4.92\\
${\rm [Ni/Fe]}${\sc i}  & 39 & 0.00   & 0.17 & 22 & 0.19   &0.10 & 6.28\\
${\rm [Ba/Fe]}${\sc ii} &  2 & 0.38   & 0.01 &  3 & 0.30   &0.01 & 2.22\\
\hline
\end{tabular}
}
\label{t:abutot}
\end{center}
\end{table}

\section{Abundances}

The iron abundances\footnote{In the analysis we used the reference solar
values given by Gratton et al. (2003) in their Table 8, col. 5; for elements
not studied in that paper (such as Co and Ba) we adopted the solar abundance
from meteorites given in Anders and Grevesse (1989). All values
are listed in Table \ref{t:abutot}.} 
are given in Tables
\ref{t:atmpar} and \ref{t:abutot} for each star, and an average of [Fe/H]{\sc
i}=$-0.08 \pm 0.05$ dex (rms=0.11, 6 stars) is found. However, this value
includes star 17814 near the RGB tip, for which the analysis
might be somewhat less reliable. In the following, we will
exclude this star when reporting the average abundance ratios. 
Therefore, we adopt a final value of [Fe/H]$=-0.03 \pm 0.03$ dex
(rms=0.03 dex, 5 stars) for the
metallicity of Cr 261.

The error quoted here is given by the scatter in the results for individual stars,
so it represents a measure of the internal errors. It agrees with (actually, it
is smaller than) the errors expected from uncertainties in the atmospheric
parameters. We do not attach much weight to this difference, since the sample
includes only 5 stars. Systematic errors, not included in these estimates, are
much larger and uncertain. However, our purpose is to construct a large and
homogeneous set of abundances for open clusters. The error bar that we quote is
then to be considered as the individual error for Cr 261 around the mean
relation defined in our series of papers.

We also derived abundance ratios [X/Fe] for the light metals (O, Na and Al),
$\alpha-$elements (Mg, Si, Ca, Ti {\sc i} and Ti {\sc ii}), elements of the
Fe-group (Sc, Cr {\sc i}, Cr {\sc ii}, Mn, Co, Ni) and the neutron-capture
element Ba {\sc ii}. Detailed abundance ratios for these elements are given in
Table \ref{t:abutot}, along with the number of lines used in the analysis and
the 1 $\sigma$ rms variance.

Mean abundance ratios for the cluster
are given in the second column of Table \ref{t:abumean}. The 1
$\sigma$ errors around the  mean values are generally comparable to, or
smaller than, what is expected  from the errors in atmospheric parameters and
$EW$s (see Table \ref{t:sensitivity}). This suggests that no intrinsic 
star-to-star scatter is present within our sample.

In the third column of Table 3 we list for comparison the original values
obtained by F03, while in the last column we recomputed their average values
by accounting for the different adopted solar abundances (in F03 they mostly
come from Anders and Grevesse 1989, apart from the iron solar value mentioned
above).

\begin{table}
\begin{center}
\caption{Elements measured and mean cluster abundance ratios with rms (second
and third column). Only 5 stars are considered. The Fe abundance is
referred to H, and we give only Fe {\sc i} because Fe {\sc ii} was forced to
agree with it. The fourth and fifth columns give the corresponding values in
F03, while the last column shows the F03 abundances converted to our adopted
solar values. 
For C and O "a" and "b" have the same meaning as in Table \ref{t:abutot}.}
\begin{tabular}{lrcrcc}
\hline
 \multicolumn{1}{c}{Element} 
&\multicolumn{1}{c}{[X/Fe]}
&\multicolumn{1}{c}{ $\sigma$ }
&\multicolumn{1}{c}{[X/Fe]}
&\multicolumn{1}{c}{ $\sigma$ } 
&\multicolumn{1}{c}{[X/Fe]}\\
 \multicolumn{1}{c}{} 
&\multicolumn{1}{c}{us}
&\multicolumn{1}{c}{ us }
&\multicolumn{1}{c}{ F03}
&\multicolumn{1}{c}{ F03 } 
&\multicolumn{1}{c}{F03,conv.}\\
\hline
Fe {\sc i}   & $-$0.03 & 0.04 &  $-$0.22 & 0.11 & $-$0.24\\
C  {\sc i} a & $-$0.11 & 0.11 &          &      &  \\
O  {\sc i} a & $-$0.12 & 0.09 &  $-$0.10 & 0.15 &  ~~0.01\\
O  {\sc i} b & $-$0.30 & 0.04 &  $-$0.10 & 0.15 &  ~~0.01\\
Na {\sc i}   &	 0.33  & 0.06 &     0.48 & 0.22 &  ~~0.62\\
Mg {\sc i}   &	 0.17  & 0.07 &     0.07 & 0.12 &  ~~0.24\\
Al {\sc i}   &	 0.12  & 0.08 &     0.39 & 0.12 &  ~~0.65\\
Si {\sc i}   &	 0.24  & 0.05 &     0.22 & 0.09 &  ~~0.26\\
Ca {\sc i}   &	 0.01  & 0.05 &  $-$0.04 & 0.10 &  ~~0.07\\
Sc {\sc ii}  &    0.12 & 0.04 &          &      &	\\ 
Ti {\sc i}   & $-$0.12 & 0.09 &  $-$0.07 & 0.09 & $-$0.06\\
Ti {\sc ii}  & $-$0.06 & 0.04 &  $-$0.06 & 0.01 & $-$0.17\\
Cr {\sc i}   & $-$0.01 & 0.06 &  $-$0.19 & 0.13 & $-$0.17\\
Cr {\sc ii}  &    0.06 & 0.09 &          &      &	\\
Mn {\sc i}   & $-$0.03 & 0.04 &          &      &	\\ 
Co {\sc i}   &	 0.01 & 0.10  &          &      &	\\ 
Ni {\sc i}   &	 0.06 & 0.08  &     0.02 & 0.04 &  ~~0.01\\
Ba {\sc ii}  &   0.30 & 0.08  &          &      &	\\
\hline
\end{tabular}
\label{t:abumean}
\end{center}
\end{table}

\subsection{The light elements O, Na, Al}

The oxygen abundances given in Table \ref{t:abutot} are only preliminary values
computed from the measured equivalent widths of the forbidden [O {\sc i}]
6300.31~\AA\  line, which is a  primary O indicator. We checked that 
the oxygen lines  in all Cr 261 stars were not
affected by telluric components by comparison to an early type star. 

Recently, Johansson et al. (2003) recomputed the laboratory oscillator
strength of the weak, high excitation Ni {\sc i} line at 6300.35~\AA, 
claimed to be a significant contaminant of the
forbidden [O {\sc i}] line  in the Sun by
Allende Prieto et al. (2001).
To evaluate this contamination, we computed the $EW$s of the Ni line 
using the [Ni/Fe] ratios we determined for our program stars in Cr 261, and the
$\log gf$ by Johansson et al. The contribution of
Ni to the $EW$s of the forbidden [O {\sc i}] line ranges from
about  20\% up to 40\% in the program stars.  

In order to derive reliable O abundances, a careful synthesis of the [O {\sc
i}] lines is necessary, including not only the contribution of the nearby Ni
line but also the coupling with C abundances. This requires solving the set
of related dissociation equations and will be deferred to a forthcoming paper
on C, N, O abundances in OCs. First estimates of the C abundances from the
spectral synthesis of the C$_2$ molecular features at 5086~\AA\ provide an
average ratio [C/Fe]$=-0.11$ dex, $\sigma=0.11$ dex (5 stars, excluding the
tip giant). With this slight underabundance of C, the average [O/Fe] ratio
from the synthesis of the [O {\sc i}] 6300.31 line would be of $-$0.12 dex
($\sigma=0.09$ dex, 5 stars). This is in good agreement with the results of
F03, who adopted the C abundance from Arcturus ([C/Fe]$=-0.06$ dex).
The derived preliminary values are shown in Table \ref{t:abutot}.

We measured the $EW$s of the Na {\sc i} doublets at 5682-88, 6154-60
and 8184-93~\AA; the Na abundaces were computed  excluding the 5682~\AA\ line,
since it is clearly discrepant. 
Our Na abundances were corrected for departures
from the LTE following Gratton et al. (1999). Abundances derived from the 5
lines in the three different doublets are in good agreement
with each other, while in F03 there
is   a larger line-to-line scatter, likely because they did not consider non-LTE
effects for the 5682-5688 \AA ~doublet.

As usual among old OCs, [Na/Fe] ratios are enhanced with respect to stars of
similar metallicity in the Galactic field (see e.g. the compilation given by
F03 in their table 7). Our average [Na/Fe] is +0.33 dex, less than that found
by F03 (they derived +0.48, or +0.45 if we exclude the RGB tip star). 
If we account for the different  solar abundances adopted, the F03 value
becomes even larger ([Na/Fe]=+0.62 dex).

Our analysis shows a very small star-to-star scatter, and this result is quite
robust, since it is based on 5 lines of 3 different doublets, falling in different
spectral regions. Apart from the slightly discrepant star 2268 (with
[Na/Fe]=+0.22), the  other four stars share the same Na abundances within a few
hundredths of dex ($<$[Na/Fe]$>$ = +0.36, rms=0.02). Although the sample is
limited, this may imply that open  cluster stars do not experience the same
phenomenon of different  star-to-star Na-enhancement affecting most globular
cluster giants (for a recent review, see Gratton et al. 2004).

For Al, we measured all the 3 subordinate doublets used by Carretta et al.
(2004b). As discussed there the correction for non-LTE effects should increase
with decreasing metallicity, therefore we consider it negligible in our 
case.

The [Al/Fe] ratios we determined are slightly above solar, but also in this 
case less than in F03's study: our average [Al/Fe] is +0.12 dex, as compared to
their value of +0.40 dex (or +0.65 dex, once corrected to our scale). With our
determination, Cr 261 does show the same behaviour - at least for Al - of field
disk stars of solar metallicity, as is the case for most old open
clusters.  So, while it appears that the [Na/Fe] ratio is enhanced (and also
shows a rather large scatter, of more than 0.4 dex) in old open clusters with
respect to their field counterparts of similar metallicity, the Al enhancement
is not so widespread.

In the past several authors have discussed the possibility that the large
enhancements in Na and (to a lesser extent) in Al  may be due to the same
mechanism acting in more metal-poor evolved giants, such as those of globular
clusters: internal mixing. However, this hypothesis
presents several shortcomings:
\begin{itemize}
\item A Na-O anticorrelation, signature of proton-capture reactions at high
temperature in the same site where the ON and NeNa cycles occur (see Gratton et
al. 2004 for a recent review), has not been observed yet in open cluster
stars.
\item it is now acknowledged that the bulk of abundance anomalies in
globular cluster stars is produced elsewhere than in the observed stars
themselves, since
star-to-star variations in all the main elements involved (C, N, O, Na, Al:
Gratton et al. 2001, Carretta et al. 2004c) are observed down to unevolved
dwarfs. Intermediate-mass AGB stars of a previous generation
are the favoured candidate polluters (see Gratton et al. 2004).  
\item It has been suggested (e.g. Lotz \& Friel 1995) that most red giants in
old open clusters exhibit little or no variation in the CN band strengths, at
odds with what is observed in globular cluster giants. In the internal mixing
scenario, C and N are the first elements to be affected by any little 
composition change due to non-canonical mechanisms connecting the outer envelope
with inner regions near the energy generating CNO-burning shell. However, no
inhomogeneities in the N abundance are detected among dwarf stars in the old
open cluster M 67 (Hufnagel 1996), and the spread in CN observed in giants of
the old cluster NGC 6791 (Hufnagel et al. 1995) is compatible with the normal
changes undergone by a star when evolving along the red giant
branch.
\end{itemize}

\subsection{The $\alpha$-elements}

We derived the abundances of the $\alpha$-elements Mg {\sc i}, Si {\sc i}, Ca {\sc
i}, Ti {\sc i} and Ti {\sc ii}. These four "$\alpha$-elements", which are
often taken together, do not appear to have all the same behaviour,  as
already noticed for other open clusters and stellar populations. 
Perhaps bundling them together is not the best way to examine the evolution of 
elements with time, or with iron abundance. 
For example, Ti can be considered 
either as the heaviest element synthesized by $\alpha$-capture processes or one
of the lightest elements in the Fe-group (e.g. Wheeler et al. 1989).

Our [Mg/Fe] and [Si/Fe] abundance ratios are both above solar  (at +0.17 dex
and +0.24 dex, respectively) for this near solar metallicity cluster.  As far
as Mg is  concerned, Cr 261 closely follows the behaviour of all the other old
OCs studied so far, and the trend with   metallicity defined by
Galactic field stars. However, Si is much more discrepant with respect to the
same field stars, both in our analysis and in the one by F03, and in about
half the clusters for which reliable Si abundances from high-resolution
spectroscopy exist.

Before giving too much weight to these results, we have to verify whether they
could be due to some error in the abundance analysis. For instance, 
a  \teff ~underestimated by about 200 K would bring the [Mg/Fe] and [Si/Fe] 
ratios
closer to solar. The use of higher temperatures would also
increase the $\log g$ values derived from the ionization equilibrium of Fe
by about 0.6 dex, putting them in better agreement with the ones 
derived from the location of the stars in the CMD.
However, such warmer temperatures would also
result in metallicities higher  by about 0.15 dex, values that are not
supported by the synthetic CMDs or by any other spectroscopic
metallicity estimate.

A more interesting possibility is that the adopted model atmospheres are not
fully adequate even for these relatively warmer stars. Temperature gradients
steeper than those of the Kurucz model atmospheres may reconcile most of the
residual discrepancies:  ionization equilibrium with gravities from the CMD;
abundance ratios of Mg and Si with Fe against solar abundances; overall metal
abundance from spectroscopy and photometry. Such steeper gradients 
might be due
to e.g. the effects of adiabatic expansion of raising gas columns in 3-d model
atmospheres, as shown by Asplund and collaborators (e.g. Asplund et al. 2000).
However, reliable 3-d model
atmospheres are not yet available for red giants. We then prefer to keep the
uniformity of our analysis throughout this series of papers, and not apply
arbitrary modifications to the standard 1-d model atmospheres. The reader
should however be cautioned  of the possibility that small systematic deviations
of abundance ratios from solar over the whole set of data might be due to the
analysis method rather than to real abundance anomalies.

Cr 261 shows a nearly solar [Ca/Fe], in agreement with the other old OCs, where
the Ca abundance traces that of Fe, although with a rather large scatter
(of the order of 0.3 to 0.4 dex). The [Ti/Fe] ratios are slightly below solar
and they closely follow the trend defined by field
stars, again as found for the other old OCs. 
Cr 261 does not stand out in these distributions, although both our
determination and F03's are located near the bottom envelope of the [Ca/Fe] vs
[Fe/H] scatter.

Finally, we note that the Ti {\sc i} and {\sc ii} values are in good agreement
with each other, supporting the surface gravities derived from the
iron lines. Note
however that the regions in the atmosphere where Ti lines form are quite
similar to those where Fe lines form (the same is valid for the Cr lines
mentioned in the next subsection), so that this test does not exclude a
steeper temperature gradient.
 
\subsection{Elements of the Fe-group and Barium}

Among the elements of the Fe-peak, the abundances of Sc {\sc ii},  Mn {\sc i}
and  Co {\sc i} include corrections for hyperfine structure (HFS, see Gratton
et al. 2003 for references).   The good agreement of the mean abundance
of Cr {\sc i} and Cr {\sc ii} supports the atmospheric parameters derived
from our full spectroscopic analysis.

Among these elements, only the [Sc/Fe]{\sc ii} ratio is slightly overabundant
with respect to the solar value, but this is the typical behaviour of   both
disk field stars and of the other old OCs. The 
ratios of Cr (from both neutral and singly ionized lines), Mn, Co, and Ni 
to iron closely trace the behaviour of solar metallicity stars with [Fe/H].

The abundances of Ba, which is produced in neutron capture
reactions, are important since they reflect the contribution to the chemical
enrichment by AGB stars.
At solar metallicity, Ba is mostly produced by the
main $s$-processes
in low mass (1-3 M$_\odot$) AGB stars, with a small contribution ($\sim$ 15\%)
from $r$-processes. The derivation of abundances in different metallicity
regimes is a good method to study the metallicity dependence of the 
nucleosynthetic yields.

Abundances of Ba were derived in all the stars of our sample. We used the two
subordinate lines of Ba {\sc ii} at 5853 and 6496~\AA\ that are free of HFS
effects (Mashonkina and Gehren 2001). In order to check for possible effects of
departures from LTE, we compared the abundances from the 5853~\AA\ line (for
which the effects are small, according to Mashonkina and Gehren) to the
abundances obtained from the other subordinate line, for which they found more
significative effects. The average difference is 
log n(Ba)$_{6496}$ - log n(Ba)$_{5853} = 0.04 \pm 0.02$, $\sigma=0.06$ dex 
(6 stars), therefore we do not expect remarkable corrections for non-LTE, in this 
case.

The average [Ba/Fe] ratio derived in Cr 261 is quite high, with small 
star-to-star scatter, well within the uncertainties of the adopted atmospheric
parameters. This would suggest a larger contribution of the low mass
producers of $s-$process elements with respect to type Ia SNe, producing Fe
but not Ba. However, at solar metallicity the [Ba/Fe] ratio in old open
clusters seems to show a large scatter (about 0.5 dex). Although this may
indicate a possible range in the ratio of contributions from AGB stars and SNe,
we caution that the many differences involved in the abundance analysis of
this element (HFS, application or lack of correction for departures from LTE)
might introduce some spurious effect, masked as an intrinsic scatter.

\section{Summary and conclusions}

In this paper we have derived atmospheric parameters and elemental abundances
for 6 giant stars in Cr 261. The iron abundance deduced from the 5 warmer stars
in our sample is [Fe/H] = $-0.03$ (rms 0.03). 

This has been compared to the values of [Fe/H]=$-0.16$ (Friel et al. 2002),
and $-$0.20 (F03), and to the value found from photometric data with the
synthetic CMD technique. Gozzoli et al. (1996) found that the best fits were
generally obtained with the solar abundance tracks, and this has been
confirmed by the new analysis with other, newer evolutionary tracks
(Bragaglia \& Tosi 2005, and in preparation): the cluster appears to be at a
distance of about 2.7 kpc from the Sun,  
to have a reddening of \ebv \ around 0.30, 
an age of about 6 Gyr and solar
metallicity.

We also measured the reddening, comparing intrinsic colours derived from the
spectroscopic temperatures and the Alonso et al. (1999) \teff-colour relations
with the observed ones. We found values compatible with those
 based purely on photometric data, but two stars show hints of possible 
contaminations  in the optical/IR photometry. We also found indication
of a possible non-standard ratio
between selective and total absorption.

We measured abundances for several elements other than iron, namely Na, Mg,
Al, Si, Ca, Sc, Ti, Cr, Mn, Co, Ni, Ba. This is important if we want to
compare the nucleosynthetic history of open clusters to that of field thin
and thick disk stars, or to the halo components.

Systematic differences should be taken into account when comparing literature
results; however, the sources of these differences are various, and a more
detailed and critical analysis is deferred to another paper. To eliminate at least part of
these systematic differences, the original  abundance ratios of several other
OCs and field disk stars were transformed to our scale by correcting for
the different solar values adopted (Bragaglia et al. in preparation). We have
compared in this way our abundances for  Cr 261 with those
previously obtained for this cluster and  with those of several other OCs and of
field disk stars. While Cr 261 seems to generally fit in the trends defined by
old open clusters, more definitive  conclusions on the
similarities/dissimilarities between open clusters and other components of the
Galaxy will have to wait until a larger sample of OCs is 
studied in a truly homogeneous way.

\begin{acknowledgements}
{We thank E. Friel for  
useful discussions and crucial information on cluster membership, and all the
authors who sent us references on their adopted solar values. E.C. 
warmly thanks Jennifer Sobeck for enlightening discussions on Mn abundances and
other issues.  
This research has made use of the SIMBAD data base, operated at CDS,
Strasbourg, France, and of the  BDA, maintained by J.-C. Mermilliod. This
publication makes use of data products from the Two Micron All Sky Survey,
which is a joint project of the University of Massachusetts and the Infrared
Processing and Analysis Center/California Institute of Technology, funded by
the National Aeronautics and Space Administration and the National Science
Foundation. This work was partially funded by Cofin 2000-MN02241491 
and Cofin 2003-029437 by the Italian MIUR. }
\end{acknowledgements}

\end{document}